# Spectrometer-free optical hydrogen sensing based on Fano-like spatial distribution of transmission in a metal-insulator-metal plasmonic Doppler grating

Yi-Ju Chen,[†] Fan-Cheng Lin,[‡] Ankit Kumar Singh,[†] Lei Ouyang,[†, §, ⊥] and Jer-Shing Huang[†,‡,§, ∥,∆,*]

[†] Leibniz Institute of Photonic Technology, Albert-Einstein Straße 9, 07745 Jena, Germany

[‡] Department of Chemistry, National Tsing Hua University, Hsinchu 30013, Taiwan

[§] Institute of Physical Chemistry and Abbe Center of Photonics, Friedrich-Schiller-Universität Jena, Helmholtzweg 4, D-07743 Jena, Germany

[⊥] State Key Laboratory of Biogeology and Environmental Geology, Faculty of Materials Science and Chemistry, China University of Geosciences, Wuhan 430074, China

[∥] Research Center for Applied Sciences, Academia Sinica, Taipei 11529, Taiwan

[∆] Department of Electrophysics, National Yang Ming Chiao Tung University, Hsinchu 30010, Taiwan

**Abstract**

Optical nanosensors are promising for hydrogen sensing because they are small, free from spark generation, and feasible for remote optical readout. Conventional optical nanosensors require broadband excitation and spectrometers, rendering the devices bulky and complex. An alternative is spatial intensity-based optical sensing, which only requires an imaging system and a smartly designed platform to report the spatial distribution of analytical optical signals. Here, we present a spatial intensity-based hydrogen sensing platform based on Fano-like spatial distribution of the transmission in a Pd-Al$_2$O$_3$-Au metal-insulator-metal plasmonic Doppler grating (MIM-PDG). The MIM-PDG manifests the Fano-resonance as an asymmetric spatial transmission intensity profile. The absorption of hydrogen changes the spatial Fano-like transmission profiles, which can be analyzed with a "spatial" Fano-resonance model and the extracted Fano resonance parameters can be used to establish analytical calibration lines. While gratings sensitivity to hydrogen absorption are suitable for hydrogen sensing, we also found hydrogen insensitive gratings, which provide an unperturbed reference signal and may find applications in nanophotonic devices, that require a stable optical response under fluctuating hydrogen atmosphere. The MIM-PDG platform is a spectrometer-free and intensity-based optical sensor that requires only an imaging system, making it promising for cellphone-based optical sensing applications.

## 1. Introduction

Hydrogen has been widely used as a coolant and raw material in the semiconductor industry and food industry.[1] As a carbon-free energy carrier with high energy density, hydrogen is the most promising clean-burning fuel resource for the future.[2] However, its properties of odorless, colorless, high flammable, wide explosive range (4-75 Vol.%), and low spark ignition energy (0.02 mJ) also arise critical safety concerns due to potential explosion risks.[3] Safety protocol implementation and real-time monitoring are essential for its application. Sensitive, accurate, and fast leakage detection, especially for detecting hydrogen below the explosive limit (< 4 %), is critical.

Various hydrogen sensing strategies have been proposed, including micro-electromechanical,[4] catalytic,[5] thermal,[6] electrochemical,[7] mechanical,[8] optical,[9] and acoustic[10] detection schemes. The core element for all these sensing systems is the hydrogen sensitive materials, such as metal oxides ($In_2O_3$,[11] $ZnO$,[12] $TiO_2$,[13] and $W_{18}O_{49}$[14]) and hydride-forming metals (Pd,[15] Mg,[16] Y,[17] alloys[18]). Among various sensing materials, metal oxides are less applicable because they require a high working temperature of over 400 °C to maintain optimal sensitivity. Therefore, hydride-forming metals, which can work at room temperature with high sensitivity and selectivity, have been commonly used in optical hydrogen sensors. In particular, Pd has been widely used in many works because of its fully reversible hydride formation properties under ambient conditions. When Pd is exposed to hydrogen gas, hydrogen molecules dissociate into atoms, which enter the interstitial sites of Pd.[19] This induces mass and volume changes in the Pd host, as well as the change in the conductivity and dielectric properties.[20] The volume change due to the hydrogen-induced lattice expansion in Pd is the basis of mechanical type hydrogen sensors.[8] The change of conductivity and optical properties stems from the fact that electrons of the hydrogen atoms enter the *s*- and *d*-bands of Pd, leading to modification in the density of electronic states at the Fermi level and causing shifts of the energy bands. This, in turn, alters the conductivity and dielectric properties of Pd. While the change of Pd's conductivity has been used in electronic hydrogen sensors,[21] the modification in the dielectric properties promises its applications in optical hydrogen sensors.[9, 22-24]

Compared to electrical hydrogen sensors, which usually suffer from possible electric sparks, optical hydrogen sensors offer advantages of remote and contact-free readout and avoid spark generation.[25] The change of refractive index alters the spectral feature of localized surface plasmon resonance (LSPR), including the broadening, shifting, and reducing the intensity of the SPR peaks.[9] However, the hydrogen-induced plasmon shifts and intensity changes are relatively small when the hydrogen volume concentration is below 4 %. To enhance the sensitivity of SPR-based hydrogen nanosensors, various strategies have been applied, including using optical nanoantennas,[26] core-shell structure,[23, 27] alloy particles,[18, 28] and composite materials.[29] Unfortunately, the sensitivity of these optical hydrogen sensors is still limited by the broad bandwidth of the LSPR peaks. This makes the detection of peak shift difficult, and thus a high-resolution spectrometer is required for hydrogen sensing. Using Fano-like resonance of specially engineered nanostructures with a steep spectral edge can be an alternative to increase the spectral sensitivity of SPR based hydrogen sensors.[30] Shifts of the corresponding resonances result in significant intensity modulation at a fixed frequency due to the steep spectral edge.[31]

For most optical sensors, a broadband illumination source and a spectrometer are mandatory to perform spectroscopic analysis. However, for practical applications, a single-color intensity distribution that can be read out without a spectrometer is more attractive.[22, 32] In this work, a two-dimensional metal-insulator-metal plasmonic Doppler grating (MIM-PDG) has been specially designed for hydrogen sensing based on Fano resonance. The MIM-PDG consists of a Pd layer on top of an insulating $Al_2O_3$ layer deposited on a single-crystalline Au flake. Upon the absorption of hydrogen gas (0 % - 4 %), the Pd layer changes its phase to Pd-H, leading to a significant change in the refractive index and volume, further modifying the azimuthal angle-dependent transmission intensity. By plotting the calibration curve of the relative change ($Q_i$) at a specific azimuthal angle as a function of hydrogen concentration, the $H_2$ uptake can be quantitatively evaluated. Due to the interference of the MIM magnetic mode inside the Pd-$Al_2O_3$-Au cavity and the propagating surface plasmon polariton (SPP) mode on the $Al_2O_3$/Au interface, the MIM-PDG platform exhibits clear Fano-like resonance in the transmission spectrum. Since the Fano resonance varies with the grating periodicity, at a fixed wavelength, the chirped grating of MIM-PDG manifests the Fano resonance as a spatial (angular) distribution of transmission intensity with a Fano-like profile. We carefully analyze this kind of single-color "spatial" Fano profile and developed an analytical model to fit the experimental data. The change in transmission profile due to $H_2$ uptake was also analyzed. Such intensity and image-based hydrogen sensing are quantitative and spectrometer-free. It only requires a simple imaging system and image analysis software. It is thus a suitable method for sensing in microfluidic channels[33] or for cellphone-based optical gas sensing.[34]

## 2. Result and discussion

### 2.1. Design and fabrication of the PDG

The Plasmonic Doppler Grating (PDG) structure was designed to provide continuous and broadband lattice momentum for photon-to-plasmon coupling. The grating periodicity is chirped in the azimuthal direction, as demonstrated in our previous works.[35-37] The trajectory of the PDG mimics the wave fronts of a moving point source of waves that exhibits Doppler Effect. Therefore, the trajectory of the n$^{th}$ ring can be described by the radius increment ($\Delta r$, similar to the source wavelength) and the ring center shift ($d$, similar to the source velocity) as

$$(x + nd)^2 + y^2 = (n\Delta r)^2 \qquad (1)$$

Figure 1a shows the schematic of a MIM-PDG with a chirped periodicity varying continuously from $\Delta r + d$ to $\Delta r - d$ which corresponds to an azimuthal angle from $\varphi = 0°$ to $\varphi = 180°$. The relationship between the periodicity of the grating (P) and azimuthal angle ($\varphi$) is given by

$$P(\varphi) = \pm d\cos\varphi + \sqrt{(d^2\cos 2\varphi + 2\Delta r^2 - d^2)/2} \qquad (2)$$

With the knowledge of incidence angle ($\alpha$), resonance order ($m$), and the material properties ($\varepsilon_m$: metal permittivity; $n_d$: effective index of the dielectric environment) the momentum matching condition for photons and plasmons via a metallic grating can be described as,

$$\frac{2\pi}{\lambda_0} n_d \sin\alpha + \frac{2m\pi}{P} = \pm \frac{2\pi}{\lambda_0} \sqrt{\frac{\varepsilon_m \cdot n_d^2}{\varepsilon_m + n_d^2}} \qquad (3)$$

Inserting Equation 2 into Equation 3, we obtain an equation describing the vacuum wavelength ($\lambda_0$) and the light-plasmon coupling azimuthal angle ($\varphi$),

$$\lambda_0 = \frac{\pm d\cos\varphi + \sqrt{(d^2\cos 2\varphi + 2\Delta r^2 - d^2)/2}}{m} \left( \sqrt{\frac{\varepsilon_m \cdot n_d^2}{\varepsilon_m + n_d^2}} - n_d \sin\alpha \right) \qquad (4)$$

For a specific PDG operating at a fixed wavelength, the photon-to-plasmon coupling angle $\varphi$ changes with the refractive index $n_d$, showing the index sensing capability of PDGs.[36] To enhance the sensitivity, Fano-resonance which offers sharp spectral shape sensitive to the change of the background refractive index is introduced in the design. A Pd-Al$_2$O$_3$-Au MIM structure is fabricated on top of chemically synthesized gold flakes (thickness = 120 nm).[38] The reason for using MIM geometry is that they support magnetic dark modes[22] which couple with the SPP resonance and lead to Fano resonance. The Al$_2$O$_3$ layer (thickness = 100 nm) was first prepared on the surface of the flakes by electron beam evaporation. On top of the Al$_2$O$_3$ layer, a 20-nm thick Pd layer was coated by electron beam evaporation. The PDG structure was then patterned on the MIM layered structure by focused-ion beam (FIB) milling. Two sets of design parameters, $\Delta r$ and $d$, have been chosen to fabricate the MIM-PDGs. The scanning electron microscope (SEM) images of the fabricated PDGs are shown in Figures 1b and 1c. For each MIM-PDG, only one circular ring was completely milled through by FIB until the ITO layer. Rest of the circular rings were slightly milled (depth = 20 nm) to create gratings only in the Pd layer. Overall, the MIM-PDG contains more than ten circular grooves and only one circular slit for light to transmit to the side of SiO$_2$ substrate. The thickness of Al$_2$O$_3$ layer was chosen based on the results of FDTD simulations, where the transmission through the completely-cut slit shows the maximum at 633 nm when the thickness of Al$_2$O$_3$ is 100 nm (Figure S1, Supporting Information). Due to the high resistance of the Al$_2$O$_3$ layer to FIB, a funnel-like slit with a wide-open gap (about 180 nm) at the top of the slit and a narrow valley (about 27 nm) at the bottom (inset in Figure 1c) was created. In the simulations, this funnel-like geometry was approximated by an effective vertical gap. Compared to the reflection-type PDG, transmission-type PDG is advantageous for its low background and high signal-to-noise ratio because only the photons that couple into surface plasmons via the grating can transmit through the subwavelength slit and be detected on the side of SiO$_2$ substrate.

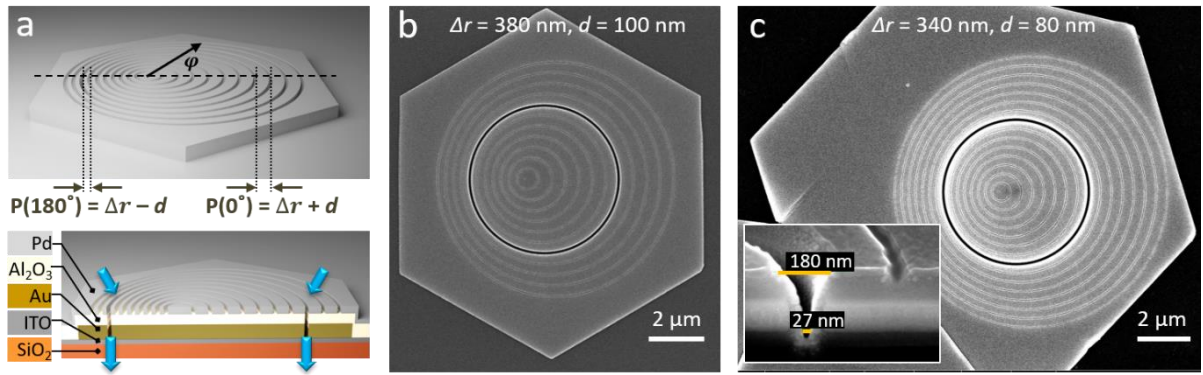

**Figure 1.** (a) Sketch showing the geometry (top panel) and cross section (bottom panel) of the Pd-Al$_2$O$_3$-Au MIM-PDG structure on an ITO coated SiO$_2$ cover slip. The grating continuously changes with the in-plane azimuthal angle ($\varphi$) from the largest period ($\Delta r + d$) at $\varphi = 0°$ to the smallest period ($\Delta r - d$) at $\varphi = 180°$. The azimuthally chirped periodicity can be freely designed by changing $d$ and $\Delta r$. (b) SEM image of a MIM-PDG fabricated with the design parameters $d = 100$ nm and $\Delta r = 380$ nm. This MIM-PDG contains ten circular grooves and one circular slit in the middle for light transmission. (c) SEM image of a MIM-PDG fabricated with the design parameters $d = 80$ nm and $\Delta r = 340$ nm. This MIM-PDG contains

fourteen circular grooves and one circular slit in the middle for light transmission. It has been optimized for single-color hydrogen sensing at 633 nm. The inset shows the zoomed-in SEM image on the cross section of the circular slit, which has a funnel-like geometry with a wide gap of 180 nm at the top surface and a narrow gap of 27 nm at the bottom.

## 2.2. Optical response of the MIM-PDG

To obtain the spectra of the transmitted light at different azimuthal angles, we used a home-built spectral mapping system (Figure 2a). Briefly, unpolarized white-light (550 - 790 nm) in the Köhler illumination scheme was blocked by a ring aperture to create ring illumination via a condenser (Air, N.A. = 0.3, Zeiss) at an incident angle of 17.45˚. Light transmitting through the slit was collected by an oil objective (Plan-Apochromat 63X oil Iris, N.A. = 0.7-1.42, Zeiss) and focused into the entrance slit of a spectrometer (SR-303i-A, Andor) by an achromatic lens (AC254-300-A-ML f = 300 mm, Thorlabs). The MIM-PDG sample was placed on a piezo stage (P-517.3CD, Physik Instrument) and scanned across the detection area defined by an effective pinhole in the detection beam path (Figure S2, Supporting Information). Figure 2b shows the image of the full-spectrum intensity obtained from the spectral mapping on the PDG shown in Figure 1b. To analyze the azimuthal angle-dependent transmission spectrum, thirteen transmission spectra were recorded along the azimuthal angle from $\varphi = 0°$ to $\varphi = 180°$ with an angle step of 15˚. The red squares in Figure 2b mark the observation areas of 3×3 pixels, from which the angle-dependent spectra were extracted. By selecting the output wavelength, transmission image at any chosen wavelength can be reconstructed. Figure 2c shows four transmission intensity images recorded at 550, 600, 650, and 700 nm. As can be seen, the intensity maxima appear at different azimuthal angles depending on the wavelength. This demonstrates the color-sorting capability of the MIM-PDG.[35] Figure 2d shows the thirteen normalized azimuthal angle-dependent spectra obtained from the experiment and FDTD simulations. The experimental spectra are well-reproduced in the simulations. The transmission spectra obtained at angles between 90° and 120° all exhibit clear Fano-like spectral line shape due to the interference of the MIM magnetic resonance with the grating-coupled delocalized SPP on $Al_2O_3$/Au interface.[39] The electromagnetic field distributions have been simulated to confirm the origin of the Fano resonance of our MIM-PDG structure (Figure S3, Supporting Information). Since the Pd grating is involved in the MIM mode, the hydrogen absorption by the Pd is expected to alter the spectral Fano-line shape of each grating and thus the spatial Fano profile of the MIM-PDG. As we will show later, this spatial Fano profile allows us to quantify hydrogen concentration by analyzing the azimuthal angle dependent intensity on MIM-PDG at a single wavelength.

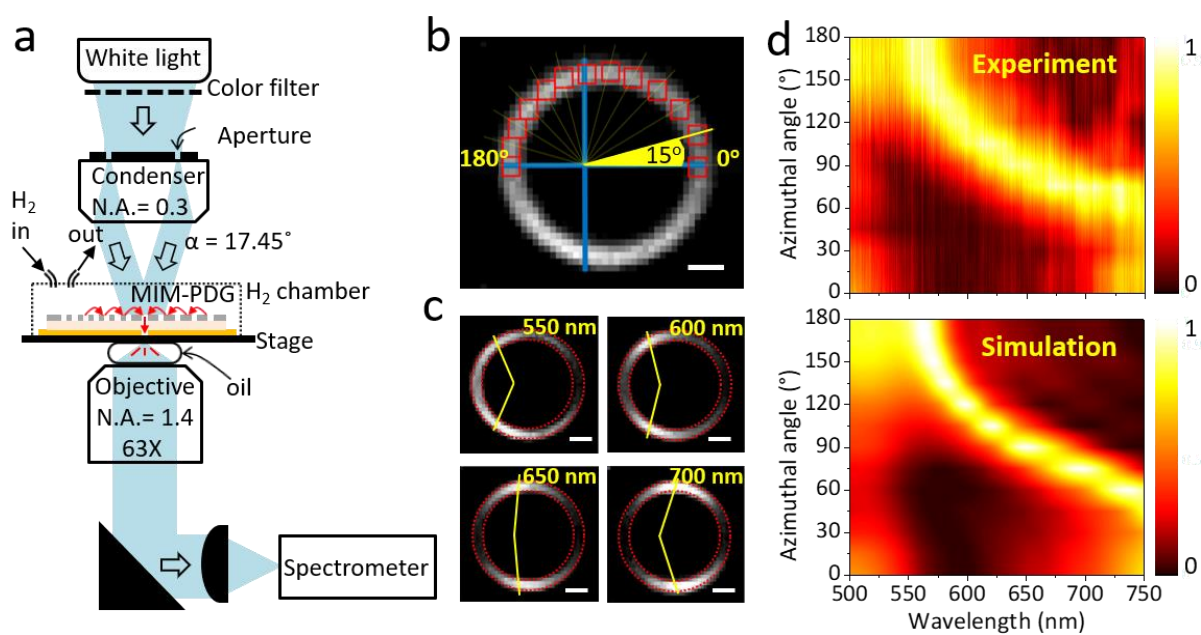

**Figure 2.** (a) Optical setup for spectral mapping on the MIM-PDG. The red arrows on the sample illustrate the propagating surface plasmons. Unpolarized ring-shaped white light illumination at α = 17.45° (N.A. = 0.3) was created by a ring aperture placed at the back focal plane of the air condenser. (b) Total transmission intensity from the circular slit of the MIM-PDG. The blue lines mark the horizontal and vertical axes of the MIM-PDG. The red squares mark the areas (3×3 pixels in the image), from which the angle-dependent spectra (angle step = 15°) were extracted. (Scale bar: 1 μm) (c) Single-wavelength transmission intensity images at four selected wavelengths, namely 550 nm (upper left), 600 nm (upper right), 650 nm (lower left), and

700 nm (lower right). (Scale bar: 1 μm) (d) Normalized transmission spectra at different azimuthal angles obtained from the experiment (upper panel) and FDTD simulations (lower panel).

### 2.3. Spectrometer-free hydrogen sensing based on Fano resonance

*Characterization of the optical response at a single wavelength*

To perform spectrometer-free hydrogen sensing, we monitored the effect of hydrogen uptake on the angle distribution of the transmission intensity at 633 nm. A similar MIM-PDG with the best optimized design parameters (d = 80 nm and Δr = 340 nm) was fabricated (Figure 1c), The optimized MIM-PDG shows a transmission maximum around $\varphi = 70°$ at 633 nm. The hydrogen gas concentration was controlled by a home-built gas mixing system as described in our previous work.[23] Briefly, the concentration of the hydrogen gas was controlled between 0 % and 4 % by tuning the ratio of the flow rate of nitrogen and hydrogen gases. The sample is placed in a home-built sample holder with a gas chamber, which allows the injection and outlet of various hydrogen-nitrogen mixtures at a constant total pressure. The overall flow rate of the nitrogen and hydrogen mixture is controlled to be 100 sccm at a total pressure of 1atm. The MIM-PDG sensor was placed in the gas chamber and the experiment was performed by illumination and detection through the transparent windows above and below the sample, respectively. The details of the gas mixing system and the sample holder can be found in our previous work.[23] Single-color illumination source was created by filtering the white light source with a laser line filter centered at 633 nm (bandwidth = 3 nm, FL632.8-3, Thorlabs). This allows us to perform hydrogen sensing by directly observing the change in the angle distribution of the transmission intensity without using a spectrometer. Different from spectral mapping described in the previous section, here, the transmitted light at 633 nm from the circular slit was directly imaged by a CCD camera (iXon, Andor, UK) with 10-second exposure time. Figure 3a shows the CCD image acquired under pure nitrogen atmosphere (0 % hydrogen). To obtain the transmission intensity angle profiles at different hydrogen concentrations in Figure 3b, we use a filter to select the slit area with a single-pixel width, each pixel corresponds to one intensity value and the pixel coordinate on the slit can be transformed into an azimuthal angle (Figure S4, Supporting Information).[35, 36] Upon the absorption of hydrogen gas, changes in the transmission intensity were observed at several specific azimuthal angles. To clearly visualize the effect on the transmission, we plotted the relative change,[22] $Q_i = (X_i-X_0)/X_0$, as a function of azimuthal angle in Figure 3c. Here, $X_0$ and $X_i$ are the angle-dependent transmission intensities measured in pure nitrogen gas (0 % hydrogen) and in a mixture gas with 1 % to 4 % hydrogen concentrations, respectively. Three features are observed in Figure 3c. First, at specific angles (e.g. $\varphi = 155°$), the deviation of the relative change from the zero line, either positive or negative, is proportional to the hydrogen concentration. This is due to the influence of the hydrogen uptake on the resonance of the Pd gratings. This allows us to easily establish a calibration line by plotting the $Q_i$ at the most sensitive period versus the hydrogen concentration. Figure 3d shows an exemplary calibration curve using the $Q_i$ from the grating period at $\varphi = 155° \pm 10°$, which is the most sensitive grating to hydrogen absorption. Secondly, the angle profile of the transmission intensity does not show angle shift upon absorption of hydrogen. This is because hydrogen uptake only slightly changes the coupling between the two modes involved in the Fano resonance. Therefore, the modification in the undulation of the Fano spectrum is more pronounced than that in the spectral shift. Consequently, the profile of the transmission angle distribution remains almost the same but the intensity undulation exhibits clearly observable changes. Thirdly, the change in the transmission intensity is not uniform but angle-dependent. Increasing hydrogen concentration can lead to an increase, decrease, or no change in the transmission intensity, depending on the azimuthal angle. For example, the transmission at $\varphi = 0°$ and $\varphi = \pm 155°$ increases with increasing hydrogen concentration, whereas the transmission at $\varphi = \pm 75°$ decreases with increasing hydrogen concentration. The transmission intensity of the grating around $\varphi = \pm 90°$ is not responding to the absorption of hydrogen. The reason for the different responses to hydrogen absorption is as follows. First, the Fano-like transmission spectrum shifts with azimuthal angles (Figure S5, Supporting Information). Therefore, for gratings at different azimuthal angles, the selected wavelength for hydrogen sensing is located at different relative spectral positions of the Fano line shape. Since the absorption of hydrogen gas only slightly changes the coupling of the two modes involved in the Fano resonance, the modification in the undulation of the Fano spectrum is much more pronounced than the spectral shift. As a result, the chosen wavelength of 633 nm can sit on the relative spectral positions that are sensitive or insensitive to the absorption of hydrogen. In other words, the variation of the spatial transmission profile depends on where the selected observation wavelength is located relative to the Fano spectrum at the specific azimuthal angle. For example, at $\varphi = \pm 75°$, the selected observation wavelength of 633 nm sits at the peak of the Fano spectrum, which decreases with the absorption of hydrogen. Differently, at $\varphi = \pm 155°$, the Fano feature is out of the observation spectral window and the selected wavelength of 633 nm is just located on a spectral position of continuum background, of which the transmission increases with hydrogen absorption. As for the gratings around $\varphi = \pm 90°$, the selected wavelength of 633 nm hits exactly the iso-point of the Fano spectrum, at which the transmission intensity is completely insensitive to the hydrogen absorption. Such angle-dependent optical response of MIM-PDG, in fact, provides very useful information for the users to select the correct grating periodicity for different targeted applications. According to Fig. 3c, gratings around $\varphi = 155°$ exhibit the largest response to hydrogen absorption and are most suitable for hydrogen sensing. If the application requires the grating to provide stable transmission insensitive to the fluctuation hydrogen concentration, the grating around $\varphi = \pm 90°$ should be used.

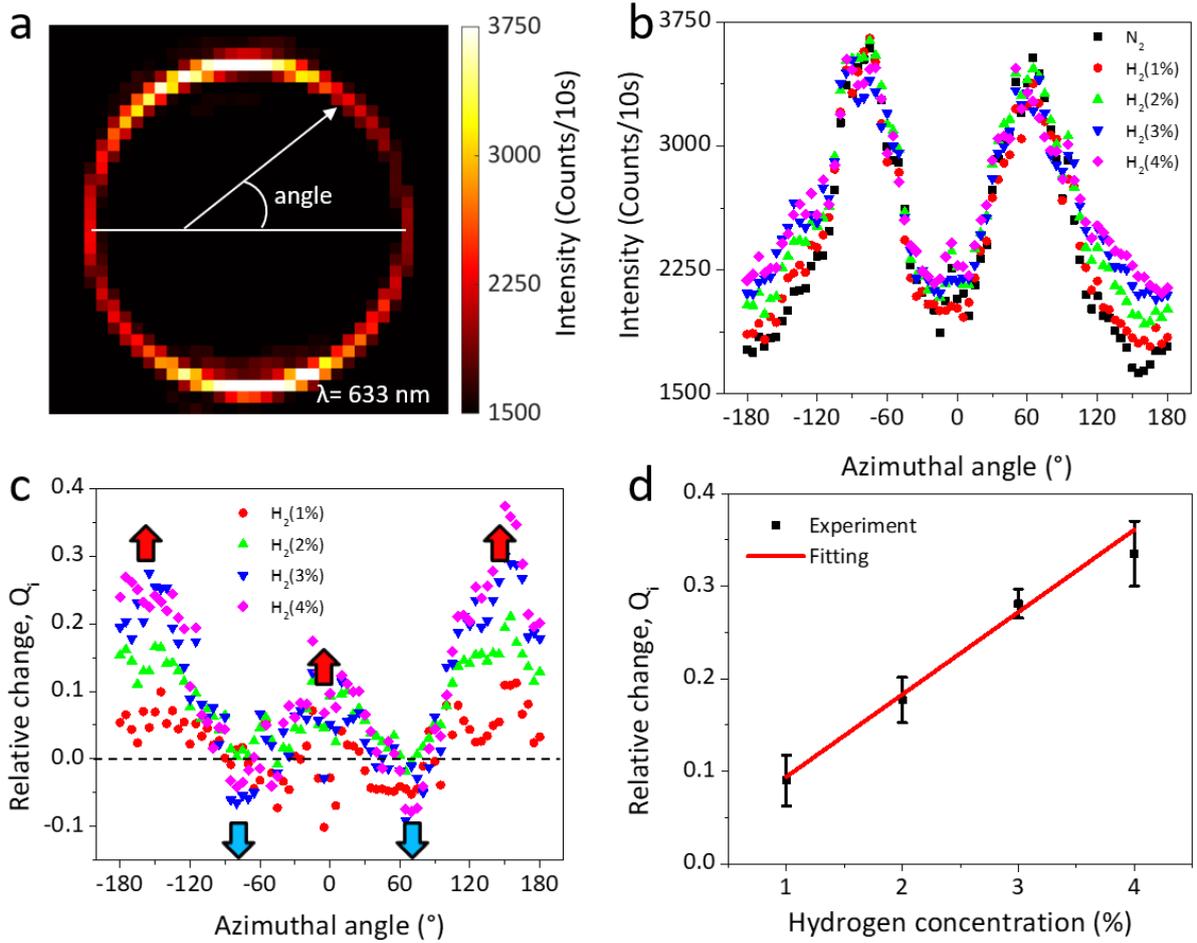

**Figure 3.** Single-color hydrogen sensing with a MIM-PDG. (a) Transmission intensity image of the MIM-PDG illuminated by an unpolarized single-color source at 633 nm illumination. The image was recorded by a CCD camera. (b) Experimental intensity angle distributions of the transmission at 633 nm measured under 0 % (pure nitrogen gas, black squares), 1 % (red dots), 2 % (green triangles), 3 % (blue inverted triangles), and 4 % (magenta diamonds) hydrogen concentrations. (c) Changes in the transmission intensity relative to 0 % hydrogen as functions of azimuthal angle for 1 % (red), 2 % (green), 3 % (blue), and 4 % (cyan) hydrogen concentrations. The red (blue) arrows mark the azimuthal angles, at which the transmission intensity increases (decreases) with increasing concentration of hydrogen. (d) Analytical calibration lines established with $Q_i$ at azimuthal angle $\varphi = 155° \pm 10°$. Transmission intensity at 633 nm was recorded by a CCD camera.

*Modelling the spatial Fano profile*

To understand the origin of the change in the angle-dependent transmission intensity, i.e. the "spatial" Fano profile, we compare the FDTD simulations with the experimentally observed changes of the intensity distribution due to the absorption of hydrogen. We investigate the change by fitting the spatial transmission profile with a spatial Fano resonance model. In our experimental system, the incident light is partially polarized after passing through various optical components in the experimental setup. This makes the angular transmission profile asymmetric about $\varphi = 0$. The elliptical polarization of the illumination can be precisely determined by examining the angular asymmetry in the transmission profile (Figure S6, Supporting Information). The parameters of the elliptical polarization were then used to correct the experimental data distortion in Figure 4a. For clarity, in the following we only show the corrected experimental results for the cases of 0 % and 4 % hydrogen concentration, representing the cases of pure Pd and Pd-H, respectively (Figures 4a). Upon the absorption of hydrogen, the experimental transmission intensity decreases and the peak width increases.

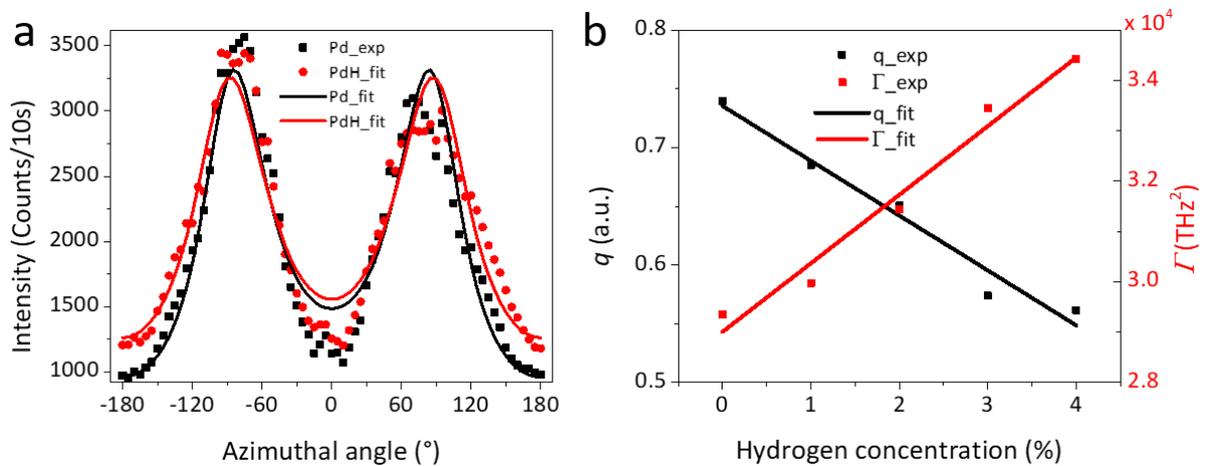

**Figure 4.** (a) Experimental angular transmission profiles recorded at 0 % (black) and 4 % (red) hydrogen concentrations and the corresponding curves from fitting the experimental data points with the spatial Fano model. (b) The q (black squares) and Γ parameters (red squares) obtained from the experimental transmission angle profiles and the corresponding calibration lines using least-square linear fit.

The spectral interference of the SPPs on the $Al_2O_3$/Au interface and the MIM mode inside the Pd-$Al_2O_3$-Au cavity gives rise to the spectral Fano resonance. Since MIM-PDG is an azimuthally chirped grating, the spectral Fano resonance of the grating at each angle can be mapped onto the "spatial" domain and the spatial (angular) profile of the transmission intensity at a fixed wavelength also exhibits Fano-like line shape. The origin of such a "spatial" Fano resonance profile along the azimuthal direction is related to the modulation of phase and amplitude relation between the MIM mode and the SPP mode. Specifically, the resonance frequency of SPP mode varies with the palladium grating periodicity along the azimuthal angle, as described by Equation 4. The resonance frequency of SPP mode excited for a given period of grating can be evaluated by the dispersion relation of surface plasmons on the $Al_2O_3$/Au interface (Figure S7, Supporting Information). The spatial Fano resonance profile along the azimuthal direction at a given frequency can be modeled by,

$$T(\varphi) = k \frac{(q+\epsilon)^2+b}{1+\epsilon^2}. \qquad (5)$$

Here, $q$ quantifies the asymmetry of the spatial Fano resonance, $b$ is the responsible for Lorentzian background associated to optical dissipation/damping, $k$ is proportionality constant and $\epsilon$ is the reduced energy scale given as,

$$\epsilon = \frac{\omega^2 - \omega_{od}^2 - \omega_{od}\Delta}{\Gamma}, \qquad (6)$$

where $\omega$ ($\frac{2\pi c}{\lambda_0}$, $c$ is the speed of light in vacuum) is the frequency of incident light, $\omega_{od}$ is the resonance frequency of the SPP mode, $\Gamma$ is the effective damping of the Fano line shape and $\Delta$ is the shift in resonant energy transfer frequency from the resonance frequency of the polariton mode. The intensity transmitted from the MIM-PDG under different hydrogen concentrations from 0 % to 4 % was fitted with Equation 5 to obtain the resonance parameters in Figure 4a. The q and $\Gamma$ parameters obtained from the fitting of azimuthal angular transmission profile are shown in Figure 4b. The numbers of all the parameters can be found in Table S1 in the Supporting Information. The parameter $\Gamma$ shows a linear increase upon absorption of hydrogen, indicating an increase of the effective damping of the Fano line shape due to the phase transition from α-phase Pd to β-phase Pd-H. The effect of hydrogen absorption is also seen in the q parameter which decreases linearly with hydrogen concentration indicating a decrease in the asymmetry of the spatial Fano resonance. The fitting parameters of the spatial Fano resonance indicate that the absorption of hydrogen leads the system towards a heavily damped Lorentzian mode. Overall, the spatial Fano profile can be fully described by our analytical model and the effect of the phase transition from Pd to Pd-H due to hydrogen absorption is correctly revealed in the fitted parameters.

## 3. Conclusions

In this work, a spectrometer-free single-color optical hydrogen sensor based on the Fano resonance in the MIM-PDG structure is demonstrated. The absorption of hydrogen results in refractive index change and volume expansion of the Pd layer, leading to a significant change of the Fano-like transmission spectrum, which further changes the spatial (angular) distribution of the transmission intensity. The MIM-PDG platform has been successfully applied to hydrogen sensing in the range of 0 % - 4 % and offers good feasibility for remote hydrogen sensing without the need for a bulky spectrometer and broadband illumination source. The MIM-PDG also provides angle-dependent optical responses of a series of gratings for the users to select the gratings highly sensitive or completely insensitive to hydrogen absorption depending on the targeted applications. Given the small footprint and spectrometer-free sensing capability, the applications of the MIM-PDG platform in lab-on-a-chip systems or cellphone-based optical sensing are anticipated.

**Corresponding Author**
*Leibniz Institute of Photonic Technology, Albert-Einstein Str. 9, 07745 Jena, Germany
E-mail: jer-shing.huang@leibniz-ipht.de

**Author Contributions**

J.-S.H. and F.-C.L. conceived the idea. F.-C.L. fabricated the structures and performed the optical measurement. Y.-J.C. performed the simulations. F.-C. L., Y.-J.C., and A.K.S. analyzed the data. All authors contributed to the preparation of the manuscript.

**Acknowledgement**

Financial support from the Thuringia State Government within its Pro Excellence initiative (APC2020), DFG (HU2626/3-1, HU2626/5-1, HU2626/6-1, SFB 1375), and the Ministry of Science and Technology of Taiwan (MOST-103-2113-M-007-004-MY3) are acknowledged.

**References**

[1] M. P. Suh, H. J. Park, T. K. Prasad, D.-W. Lim, Chemical reviews 2011, 112, 782.
[2] I. Dincer, C. Acar, International journal of hydrogen energy 2015, 40, 11094.
[3] R. Jiang, F. Qin, Q. Ruan, J. Wang, C. Jin, Advanced Functional Materials 2014, 24, 7328.
[4] M. B. Gerdroodbary, A. Anazadehsayed, A. Hassanvand, R. Moradi, International Journal of Hydrogen Energy 2018, 43, 5770.
[5] A. Harley-Trochimczyk, J. Chang, Q. Zhou, J. Dong, T. Pham, M. A. Worsley, R. Maboudian, A. Zettl, W. Mickelson, Sensors and Actuators B: Chemical 2015, 206, 399.
[6] S. Kim, Y. Song, H.-R. Lim, Y.-T. Kwon, T.-Y. Hwang, E. Song, S. Lee, Y.-I. Lee, H.-B. Cho, Y.-H. Choa, International Journal of Hydrogen Energy 2017, 42, 749.
[7] H.-W. Yoo, S.-Y. Cho, H.-J. Jeon, H.-T. Jung, Analytical chemistry 2015, 87, 1480.
[8] B. Xu, Z. Tian, J. Wang, H. Han, T. Lee, Y. Mei, Science advances 2018, 4, eaap8203.
[9] C. Wadell, S. Syrenova, C. Langhammer, ACS nano 2014, 8, 11925.
[10] D. Sil, J. Hines, U. Udeoyo, E. Borguet, ACS applied materials & interfaces 2015, 7, 5709.
[11] A. Shanmugasundaram, B. Ramireddy, P. Basak, S. V. Manorama, S. Srinath, The Journal of Physical Chemistry C 2014, 118, 6909.
[12] Z. U. Abideen, H. W. Kim, S. S. Kim, Chemical Communications 2015, 51, 15418.
[13] P. A. Russo, N. Donato, S. G. Leonardi, S. Baek, D. E. Conte, G. Neri, N. Pinna, Angewandte Chemie International Edition 2012, 51, 11053.
[14] W. Cheng, Y. Ju, P. Payamyar, D. Primc, J. Rao, C. Willa, D. Koziej, M. Niederberger, Angewandte Chemie International Edition 2015, 54, 340.
[15] J. Hong, S. Lee, J. Seo, S. Pyo, J. Kim, T. Lee, ACS applied materials & interfaces 2015, 7, 3554.
[16] T. Shegai, C. Langhammer, Advanced Materials 2011, 23, 4409.
[17] N. Strohfeldt, A. Tittl, M. Schäferling, F. Neubrech, U. Kreibig, R. Griessen, H. Giessen, Nano letters 2014, 14, 1140.
[18] C. Wadell, F. A. A. Nugroho, E. Lidström, B. Iandolo, J. B. Wagner, C. Langhammer, Nano letters 2015, 15, 3563.
[19] R. Griessen, N. Strohfeldt, H. Giessen, Nature materials 2016, 15, 311.
[20] Y. Pak, N. Lim, Y. Kumaresan, R. Lee, K. Kim, T. H. Kim, S. M. Kim, J. T. Kim, H. Lee, M. H. Ham, Advanced Materials 2015, 27, 6945.
[21] M. Cho, J. Zhu, H. Kim, K. Kang, I. Park, ACS Applied Materials & Interfaces 2019.
[22] A. Tittl, P. Mai, R. Taubert, D. Dregely, N. Liu, H. Giessen, Nano letters 2011, 11, 4366.
[23] K. C. Ng, F.-C. Lin, P.-W. Yang, Y.-C. Chuang, C.-K. Chang, A.-H. Yeh, C.-S. Kuo, C.-R. Kao, C.-C. Liu, U.-S. Jeng, Chemistry of Materials 2017, 30, 204.
[24] P. Ngene, T. Radeva, M. Slaman, R. J. Westerwaal, H. Schreuders, B. Dam, Advanced Functional Materials 2014, 24, 2374; X. She, Y. Shen, J. Wang, C. Jin, Light: Science & Applications 2019, 8, 4.

[25]	M. Matuschek, D. P. Singh, H. H. Jeong, M. Nesterov, T. Weiss, P. Fischer, F. Neubrech, N. Liu, Small 2018, 14, 1702990; X. Duan, S. Kamin, F. Sterl, H. Giessen, N. Liu, Nano letters 2016, 16, 1462.
[26]	A. Tittl, C. Kremers, J. Dorfmüller, D. N. Chigrin, H. Giessen, Optical Materials Express 2012, 2, 111; N. Liu, M. L. Tang, M. Hentschel, H. Giessen, A. P. Alivisatos, Nature materials 2011, 10, 631.
[27]	C. Y. Chiu, M. H. Huang, Angewandte Chemie International Edition 2013, 52, 12709; S. Rodal-Cedeira, V. Montes-García, L. Polavarapu, D. M. Solís, H. Heidari, A. La Porta, M. Angiola, A. Martucci, J. M. Taboada, F. Obelleiro, Chemistry of Materials 2016, 28, 9169; H. K. Yip, X. Zhu, X. Zhuo, R. Jiang, Z. Yang, J. Wang, Advanced Optical Materials 2017, 5, 1700740.
[28]	F. A. A. Nugroho, R. Eklund, S. Nilsson, C. Langhammer, Nanoscale 2018, 10, 20533.
[29]	M. Victoria, R. J. Westerwaal, B. Dam, J. L. van Mechelen, ACS Sensors 2016, 1, 222.
[30]	B. Luk'yanchuk, N. I. Zheludev, S. A. Maier, N. J. Halas, P. Nordlander, H. Giessen, C. T. Chong, Nature materials 2010, 9, 707; M. Mesch, T. Weiss, M. Schäferling, M. Hentschel, R. S. Hegde, H. Giessen, ACS sensors 2018, 3, 960.
[31]	M. Rahmani, B. Luk'yanchuk, M. Hong, Laser & Photonics Reviews 2013, 7, 329; F. Hao, Y. Sonnefraud, P. V. Dorpe, S. A. Maier, N. J. Halas, P. Nordlander, Nano letters 2008, 8, 3983.
[32]	T. Shegai, P. Johansson, C. Langhammer, M. Käll, Nano letters 2012, 12, 2464.
[33]	M. ElKabbash, K. V. Sreekanth, Y. Alapan, M. Kim, J. Cole, A. Fraiwan, T. Letsou, Y. Li, C. Guo, R. M. Sankaran, U. A. Gurkan, M. Hinczewski, G. Strangi, ACS Photonics 2019, 6, 1889; K. Yang, S. Zong, Y. Zhang, Z. Qian, Y. Liu, K. Zhu, L. Li, N. Li, Z. Wang, Y. Cui, ACS Applied Materials & Interfaces 2020, 12, 1395.
[34]	G. Rateni, P. Dario, F. Cavallo, Sensors 2017, 17, 1453; L. Scholz, A. O. Perez, B. Bierer, P. Eaksen, J. Wöllenstein, S. Palzer, IEEE Sensors Journal 2017, 17, 2889; D. Popa, F. Udrea, Sensors 2019, 19; L. E. McHale, B. Martinez, T. W. Miller, A. P. Yalin, Opt. Express 2019, 27, 20084.
[35]	K.-M. See, F.-C. Lin, J.-S. Huang, Nanoscale 2017, 9, 10811.
[36]	F.-C. Lin, K.-M. See, L. Ouyang, Y.-X. Huang, Y.-J. Chen, J. Popp, J.-S. Huang, Analytical Chemistry 2019, 91, 9382.
[37]	L. Ouyang, T. Meyer-Zedler, K.-M. See, W.-L. Chen, F.-C. Lin, D. Akimov, S. Ehtesabi, M. Richter, M. Schmitt, Y.-M. Chang, S. Gräfe, J. Popp, J.-S. Huang, ACS Nano 2021, 15, 809.
[38]	J.-S. Huang, V. Callegari, P. Geisler, C. Brüning, J. Kern, J. C. Prangsma, X. Wu, T. Feichtner, J. Ziegler, P. Weinmann, M. Kamp, A. Forchel, P. Biagioni, U. Sennhauser, B. Hecht, Nature Communications 2010, 1, 150.
[39]	A. Christ, T. Zentgraf, S. G. Tikhodeev, N. A. Gippius, J. Kuhl, H. Giessen, Physical Review B 2006, 74, 155435.

**TOC**

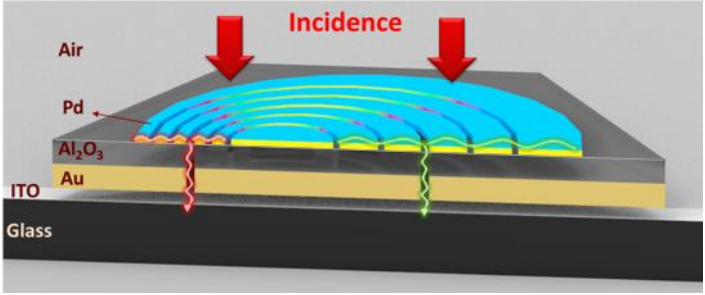

# Supplementary Information

**Spectrometer-free optical hydrogen sensing based on Fano-like spatial distribution of transmission in a metal-insulator-metal plasmonic Doppler grating**

Yi-Ju Chen,[†] Fan-Cheng Lin,[‡] Ankit Kumar Singh,[†] Lei Ouyang,[†, §, ⊥] and Jer-Shing Huang[†,‡,§, ∥, ∆,*]

[†]Leibniz Institute of Photonic Technology, Albert-Einstein Straße 9, 07745 Jena, Germany

[‡] Department of Chemistry, National Tsing Hua University, Hsinchu 30013, Taiwan

[§] Institute of Physical Chemistry and Abbe Center of Photonics, Friedrich-Schiller-Universität Jena, Helmholtzweg 4, D-07743 Jena, Germany

[⊥] State Key Laboratory of Biogeology and Environmental Geology, Faculty of Materials Science and Chemistry, China University of Geosciences, Wuhan 430074, China

[∥] Research Center for Applied Sciences, Academia Sinica, Taipei 11529, Taiwan

[∆] Department of Electrophysics, National Yang Ming Chiao Tung University, Hsinchu 30010, Taiwan

*Corresponding Author
E-mail: jer-shing.huang@leibniz-ipht.de

**Contents**
I. The transmission spectrum of MIM-PDG at different $Al_2O_3$ thickness
II. Signal acquisition in MIM-PDG measurement
III. Mode analysis for the Fano resonance
IV. Transformation of CCD images to angular intensity profiles
V. $H_2$ concentration-dependent transmission spectra of the gratings
VI. Distortion of the transmission angular profile due to partially polarized illumination
VII. Dispersion relation of the SPP mode on gold substrate
VIII. Fano parameters obtained from fitting

## I. The transmission spectrum of MIM-PDG at different Al₂O₃ thickness

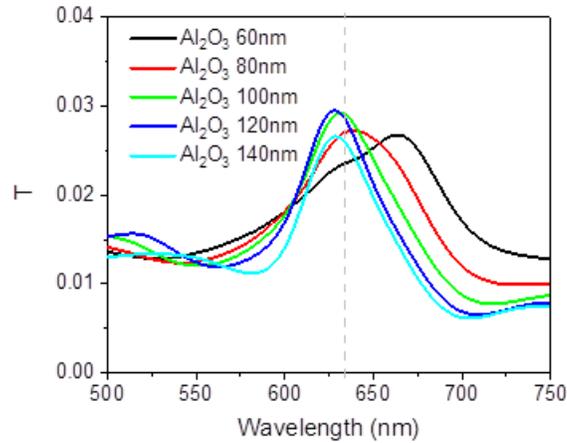

**Figure S1.** Simulated transmission spectra of the MIM-PDG obtained with various thickness of the Al₂O₃ layer. Al₂O₃ layer with a thickness of 100 nm (green trace) was found to give the highest transmission at 633 nm.

To maximize the transmission intensity at 633 nm, the thickness of Al₂O₃ has been scanned from 60 nm to 140 nm in steps of 20 nm in the FDTD simulation. According to the simulated spectrum, the transmission intensity through the completely-cut slit of the MIM-PDG reaches maximum at 633 nm when the thickness of the Al₂O₃ layer is 100 nm. The wavelength of 633 nm is indicated by the vertical dotted line.

## II. Signal acquisition in MIM-PDG measurement

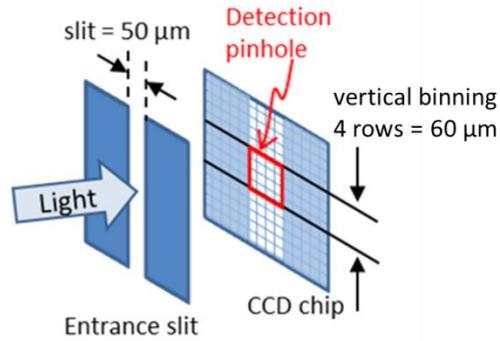

**Figure S2.** Scheme of the detection area defined by the effective pinhole.

The detection area is defined by an effective pinhole in the detection beam path. The size of the effective pinhole is around 50 μm × 60 μm, determined by the entrance slit of the spectrometer (50 μm) and the number of CCD pixels (4 pixels, 15×15 μm$^2$, Andor iDus DV412) binned vertically. Considering the magnification of the objective (63 X), the magnification of the lens system that brings the image to the entrance slit (1.82 X, from a 16.5-cm lens & a 30-cm achromatic lens), the corresponding detectable area at the sample plane is calculated to be about 437×524 nm$^2$. This provides a satisfactory spatial resolution for the spectral mapping.

## III. Mode analysis for the Fano resonance

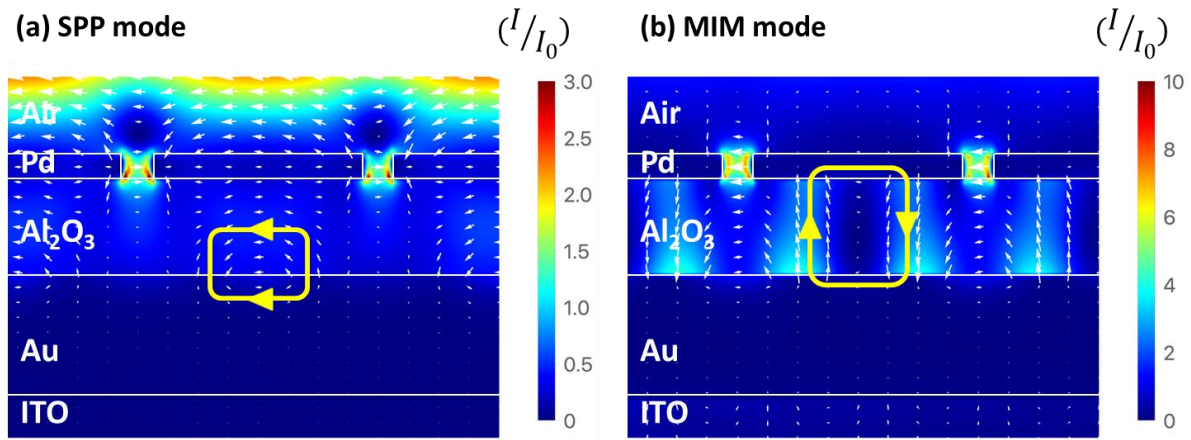

**Figure S3.** The electric field intensity distribution combined with the electric field vector plot at the resonance of the (a) SPP mode and (b) MIM mode with the grating at azimuthal angle of 75°.

For SPP mode at 554 nm (Figure S3a), the Pd grating provides a lattice momentum to the incident light to couple to the surface plasmon propagating along the interface of $Al_2O_3$ and Au. For MIM mode at 645 nm (Figure S3b), there is a strong circulating field exists in the MIM cavity. This corresponds to a magnetic mode.

## IV. Transformation of CCD images to angular intensity profiles

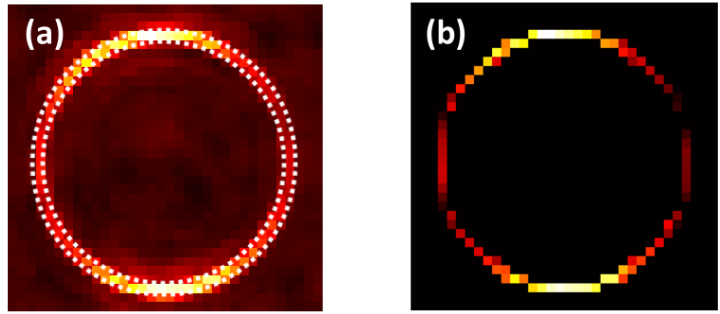

**Figure S4.** (a) The experimentally acquired CCD image and (b) the processed CCD image of the MIM-PDG

To obtain the angular transmission intensity profile, we applied a circular ring mask to the images. This mask defines a circular ring area with a width of about 200 nm width (area between the dotted circles in Figure S4a). All the background signals from the areas outside the circular ring are discarded. Only signals from the pixels inside the circular ring area are used. The ring image (Figure S4b) was then further transformed from the pixel coordinate to the corresponding azimuthal angles. In this way, the transmission intensity as a function of the azimuthal angle was plotted.

## V. H$_2$ concentration-dependent transmission spectra of the gratings

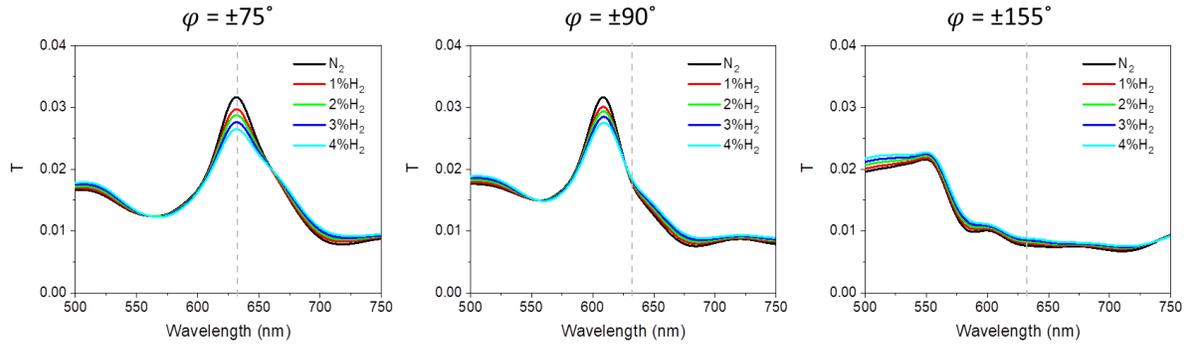

**Figure S5.** The simulated transmission spectra of the gratings at $\varphi = \pm 75°$, $\pm 90°$, and $\pm 155°$ under 0 (pure N$_2$), 1 %, 2 %, 3 %, and 4 % H$_2$ concentration.

The coupling of the MIM mode and SPP mode gives the Fano-line shape in transmission. The transmission is blue shifted with increasing azimuthal angle ($\varphi$) because the resonance wavelength of the MIM cavity corresponds to the grating periodicity. As can be seen in Figure S5, the selected observation wavelength (633 nm, grey dashed line) resides at different spectral position relative to the Fano spectrum. For gratings around $\varphi = \pm 90°$, 633 nm exactly hits the iso-point of the Fano spectrum. Therefore, the transmission is insensitive to the hydrogen uptake. Differently, for gratings around $\varphi = \pm 75°$, 633 nm just sits on the maximum of the Fano spectrum, which decreases its intensity with increasing hydrogen concentration. Therefore, the transmission from these gratings are sensitive and suitable for hydrogen sensing. For gratings at $\varphi = \pm 155°$, 633 nm is located on a continuum tail of the Fano spectrum, which increases the intensity with increasing hydrogen concentration. As a result, the transmission intensity from gratings at these angles is also sensitive and suitable for hydrogen sensing.

# VI. Distortion of the transmission angular profile due to partially polarized illumination

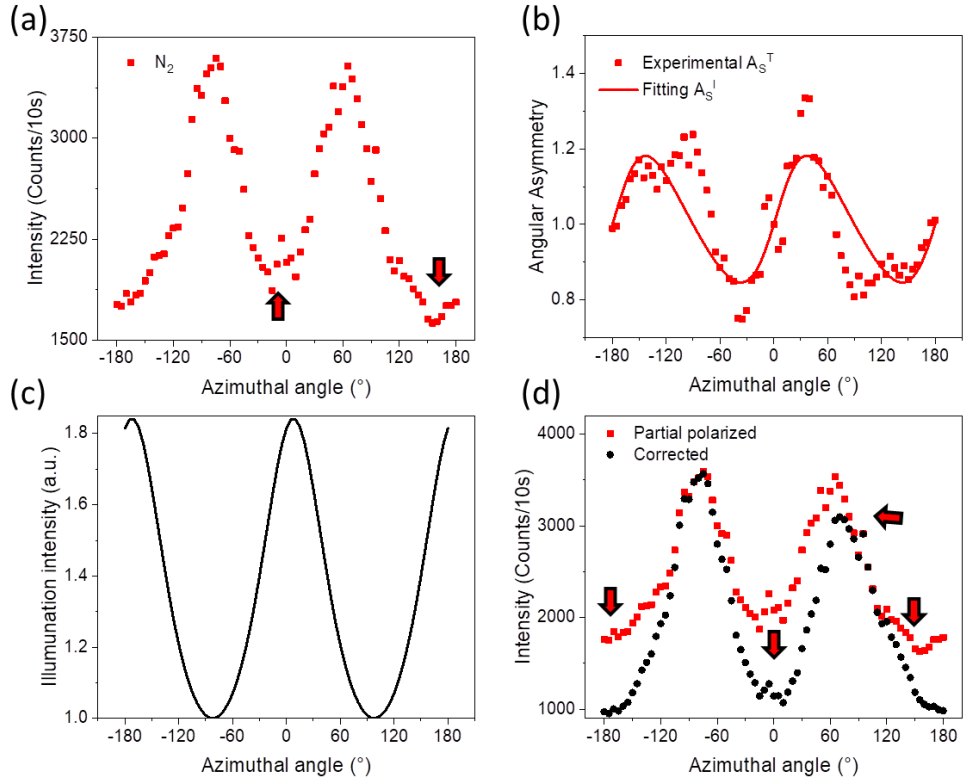

**Figure S6.** (a) The experimentally measured angular transmission profile for pure $N_2$ (0 % $H_2$). (b) The angular asymmetry, $A_s$, of the transmission profile of pure $N_2$ (0 % $H_2$) fitted with an elliptical polarization intensity asymmetry factor (major axis $a = 0.8051$ and orientation $= 97.81°$ ). (c) The azimuthal angle dependent intensity distribution of the modelled elliptically polarized source. (d) A comparison between the angular transmission profile for a partially polarized (red) and a polarization corrected (black) source. The arrows highlight the changes in angular profile due to the correction.

PDGs are designed to be illuminated by perfectly unpolarized or circularly polarized light sources because they provide equal intensity for transverse electric (TE) and transverse magnetic (TM) components present in the field at all points. Under perfectly unpolarized or circularly polarized illumination, the angular transmission intensity should always be symmetric about the axis at $\varphi = 0°$ due to the identical periodicity at $\pm\varphi$. However, in our experimental system, the incident unpolarized light from the source can be distorted by various optical components in the experimental setup, making the incident light partially polarized. This leads finite ellipticity of the polarization and results in different illumination efficiency at different azimuthal angles ($\varphi$). Consequently, the angular transmission profile is asymmetric about $\varphi = 0°$. Mathematically, the angular transmission profile ($T$) of the sample can be expressed as,

$$T(\varphi) = R_{pdg}(\varphi)I(\varphi) \tag{S1}$$

Here, $R_{pdg}(\varphi)$ is the azimuthally symmetric optical response of the sample and $I(\varphi)$ is the intensity of azimuthally asymmetric TM component of partially polarized source. The polarization corrected angular transmission profile ($T_{corrected}$) or the true response of the sample ($R_{pdg}$) is given as,

$$T_{corrected}(\varphi) = R_{pdg}(\varphi) = \frac{T(\varphi)}{I(\varphi)} \tag{S2}$$

Figure S6a shows the angular transmission profile for 0 % $H_2$ concentration with broken angular symmetry. It reveals that the efficiency of TM illumination is not symmetric about $\varphi = 0$. The asymmetric response of the MIM-PDG sample can be quantitively corrected by assuming the light incident on the sample as an elliptically polarized light. The effective parameters of the elliptically polarized source can be obtained based on the fact that the PDG's angular transmission profile is symmetric about the horizontal axis at $\varphi = 0$. Here we defined a function to quantify the angular asymmetry in the transmission profile as,

$$A_s^T(\varphi) = \frac{T(\varphi)}{T(-\varphi)} \tag{S3}$$

Since $I(\varphi)$ becomes independent of $\varphi$ for a perfectly unpolarized/circular polarized light source the $A_s^T(\varphi)$ becomes unity. Figure S6b shows a plot of $A_s^T$ with the azimuthal angle. The deviation from unity is clearly observed, confirming that the light incident on PDG has an elliptical polarization, instead of perfectly unpolarized or perfectly circularly polarized. To correct the deviation, we assume the incident light to be elliptically polarized with the major axis $a$, minor axis $b = \sqrt{1-a^2}$, and an orientation $\psi$ with $\varphi = 0$ in the anti-clockwise direction.[1] The radius of the polarization ellipse is thus given as,

$$r(\varphi) = \frac{ab}{\sqrt{a^2 \cos^2(\varphi-\psi) + b^2 \sin^2(\varphi-\psi)}} \tag{S4}$$

Now, the elliptically polarized source provides different intensity ($I(\varphi) = r^2(\varphi)$) of the TM components at different $\varphi$. The measure of angular asymmetry in the intensity profile in such case can be expressed as,

$$A_s^S(\varphi) = \frac{r^2(\varphi)}{r^2(-\varphi)} \tag{S5}$$

With the help of equations S1, S3 and S5, the intensity asymmetry of the elliptically polarized source can be directly linked to the asymmetry in the angular transmission profile of the sample as $A_s^T = A_s^S$. Figure S6b shows the fit of experimental measure of transmission asymmetry with the intensity asymmetry of elliptical polarized source (see equation S3). The fit is used to obtain the elliptically polarized source parameters that can effectively model the partially polarized light. Figure S6c shows the illumination intensity of the elliptical source ($r^2/b^2$) used to model the partially polarized source. The intensity of the elliptical polarized source is scaled by $1/b^2$ from the fitted elliptical polarization source to keep the maximum of experimental transmission profile same. The elliptical source shows a clear asymmetry in the angular intensity distribution about $\varphi = 0°$. The parameters of the elliptical source are then used to correct the experimental angular transmission profile due to partially polarized source with the help of equation S2. Figure S6d shows a comparison between the experimental angular transmission profile due to partially polarized source and the polarization corrected angular transmission profile. The apparent changes that can be seen in the corrected profile is the symmetric angular positioning of the peak transmission, narrowing of the transmission profile, and the removal of intensity increase between 160° to 180° that is obtained due to the elliptically polarized source.

## VII. Dispersion relation of the SPP mode on gold substrate

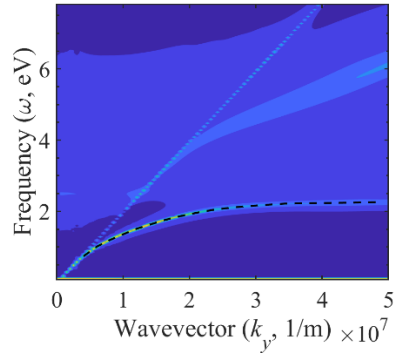

**Figure S7.** The band structure of MIM geometry obtained using FDTD simulations. The dotted black line marks the dispersion relation of the SPP mode on the $Al_2O_3$/Au interface.

The band structure of the system containing a 100-nm thick Al-coated gold substrate was simulated using the FDTD method. Figure S7 shows the band structure of grating where the intensity maximum at the lower energy marks the SPP mode of the gold substrate in the MIM geometry. The dispersion relation of the SPP mode obtained from the band structure can be used to evaluate the resonance frequency of SPP mode excited for a given period of the grating using $P = \frac{2\pi}{k_y}$.

## VIII. Fano parameters obtained from fitting

**Table S1.** Fano parameters obtained from the fitting of experimental results

|       | Effective damping of the Fano line shape, $\Gamma(THz^2)$ | Asymmetry of the Fano resonance (q) | Lorentzian background associated to optical dissipation/damping (b) |
|---|---|---|---|
| $N_2$     | 29356.14 | 0.7394 | 12.61 |
| $H_2$ (1 %) | 29975.89 | 0.6851 | 12.8  |
| $H_2$ (2 %) | 31437.56 | 0.6504 | 13    |
| $H_2$ (3 %) | 33454.66 | 0.5738 | 13.1  |
| $H_2$ (4 %) | 34431.05 | 0.5615 | 13.9  |

## References

[1]    D. H. Goldstein, *Polarized light*. Boca Raton, Fla: CRC Press. **2011**